\documentclass[preprint,12pt]{elsarticle}

\usepackage{amssymb}

\usepackage{hyperref}
\usepackage{natbib}

\hypersetup{
    colorlinks=true,
    linkcolor=blue,
    filecolor=gray,
    urlcolor=blue,
    citecolor=blue,
}

\usepackage{graphicx} 
  \usepackage{epstopdf}
  \usepackage{amsmath}
\usepackage{amssymb}
\makeatletter
\renewcommand\@biblabel[1]{}

\makeatother

\begin{document}

\begin{frontmatter}



\title{The optimal investment strategy of a DC pension plan  under deposit loan spread and the O-U process}


\author[1]{Xiao Xu }
\author[2]{Yonggui Kao \corref{cor1}}
\cortext[cor1]{Corresponding author: Yonggui Kao }
\address[1]{Department of Mathematics, Harbin Institute of
Technology, Weihai, 264209, PR China}
\address[2]{Department of Mathematics, Harbin Institute of
Technology, Weihai, 264209, PR China}


\normalsize \makeatother

\begin{abstract}

This paper is devoted to invest an optimal investment strategy for a defined-contribution (DC) pension
plan under the Ornstein-Uhlenbeck  (O-U) process and the loan. By  considering risk-free asset, a risky asset driven by O-U process  and  a loan in the financial market, we firstly set up the dynamic equation and the asset market model
 which  are instrumental in achieving the expected utility of ultimate wealth at retirement. Secondly, the corresponding Hamilton-Jacobi-Bellman(HJB) equation   is derived by means of dynamic programming principle.
The explicit expression for the optimal investment strategy is obtained  by Legendre
transform method. Finally, different parameters are selected to simulate the explicit solution and the financial interpretation of the optimal investment strategy is given.

\end{abstract}




\end{frontmatter}


\section{Introduction}

With the development of the global economy and society, pension is getting more important for the life of the elder. Besides,
nowadays the aging of population is accelerating rapidly, and pension has become a focus. The enterprise annuity is divided into two basic modes:  The defined benefit (DB) plan and the defined
contribution (DC) plan. In the DC pension plan, it transfers the longevity and financial risks from the sponsor to the
member and the DC pension plan is also playing an  role in social security, which can not be ignored. Hence, asset allocation strategy is crucial to the distribution and deployment of DC pension funds.

In recent years, many scholars have focused on the optimal investment performance.
{\citet{MarkowitzPORTFOLIO}} put forward the optimal portfolio problem for the first time and gave a theoretical proof.
Interest rate was proposed by {\citet{Duffie1996A}}  who described the general affine process.
{\citet{BoulierOptimal}}  studied the asset allocation problem of DC type enterprise fund where the interest rate obeys the the framework of Vasicek, and obtained the analytic solution through the use of martingale method.
 The expected utility directly after the retirement pension was used in the paper of {\citet{Blake2003Pensionmetrics}} .
Through the expected utility maximization, {\citet{CHARUPAT2002199}} found fixed and variable instantaneous annuities in the optimal combination of the different assumptions about mortality rates, and then made a comparison  with the optimal situation of the cumulative period.
 {\citet{Devolder2003Stochastic}} assumed that the price process of risky assets are networked with geometric Brownian motion(GBM), that is, the price fluctuation of risky assets is set as constant.
 {\citet{BaevOn}} introduced O-U process instead of GBM.
Moreover, inflation risk  on the optimal DC pension was considered in {\citet{BattocchioOptimal}}.
 {\citet{GerrardOptimal}} concentrated on income by using the stochastic optimal control technology.
Afterwards, {\citet{JianwuThe}}  obtained an explicit solution by applying the CEV model.
{\citet{GaoStochastic}} examined the complete financial market with stochastic evolution of interest rate, and used Legendre transform to settle the optimal asset allocation strategy of DC type enterprise fund.
{\citet{Hsu2008Constant}} used the CEV model for asset pricing formulas.
{\citet{GU2010580}} considered  the optimal reinsurance and investment problem of Brownian motion with risk pricing process, and the assets are described by the constant elastic variance model.
 The CRRA utility maximization and mean-variance criteria were employed by {\citet{HanOptimal}  to determine DC plan.
{\citet{Optimal}} further paid attention to the optimal allocation of DC pension with random wage under affine interest rate model.
 {\citet{GuanOptimal} studied the optimal allocation of DC pension under the framework of random interest rate and random fluctuation, in which the interest rate obeys an affine interest rate structure.
{\citet{Teng2016ON} came up with the interest rate which is subject to O-U process.
{\citet{Sun2017The}  proposed the  expected investment goal  based on deficit and surplus.
{\citet{Tang2018Asset} made an on-the-spot investigation with two situations: random interest rate and annuity inflation.
 The optimal allocation scheme with stochastic interest rate and stochastic volatility was characterised by {\citet{WangRobust}.
{\citet{BIAN201878} paid more attention to a discrete-time model with mean-variance  by a Markov chain.
{\citet{unknown} further  stated  a investment problem, which consists of  inflation and mortality risks.
Optimal investment with transaction cost over an infinite horizon was developed by {\citet{BelakFinite}.
 Based on previous work, {\citet{Mudzimbabwe}}  investigated a unsophisticated numerical solution method.  {\citet{ZhipingChen} construsted the investmal  strategy for fund administers  in a framework of Markov. A jump diffusion model was demonstrated by {\citet{Walter} . {\citet{article} attempted to apply  S-shaped utility. According to {\citet{ZHANG2020112536}, mean-variance criterion and the Cox-Ingersoll-Ross (CIR) model were adopted.
 Most of the above literatures are involved with  (CIR) model, Vasicek model, variance model and etc, however few of them apply O-U model. At the same time, they do not take loan into account in their financial market. We know that it is more accurate to adopt O-U process which reflects the fluctuation of asset price. What's more, with the upgrading and adjustment of China's industry, capital driven by economic development will become the main thrust, that is to say, the era of capital economy has come when loan will be a normal state.

Based on the above settings,
a risky asset is  depicted by the O-U process in this paper. In the framework of a discrete-time,
the business administrator is to make the expectation of the terminal wealth under the  utility framework before the retirement.
By adopting the theory of the stochastic control, the
original nonlinear HJB equation is achieved which is hard to depict closed-form expressions. And then, we introduce the Legendre transform
and separation of variables. In this case, nonlinear partial differential equations(NPDE) are transformed into linear partial differential equations.
Finally, we derive the explicit expressions of the DC scheme.
In summary, this article has two innovations: (i) we describe  the optimal investment
problem under the O-U process with CRRA utility function;
(ii) Deposit and loan spreads are taken into account, and related financial explanation is presented.

The rest is laid out as bellow. Section 2  characterizes the assumptions of the model.
Section 3  shows the definition of the value function and  derives the corresponding HJB equations and by using principle of dynamic programming.
Section 4 completes the closed-form solutions for the stochastic dynamic programming problem  under Legendre transform and CRRA utility function.
In section 5, we present the numerical simulation analysis.
We have made a summary in the final chapter.

\section{The economy and model}
\par
We  list the following assumptions for our model.

\textbf{Assumption 1.} Consider a financial market which ignores  transaction fees. We use
a finite-time horizon and continuous-time model. The uncertainty is
represented by a complete probability space ~$\left({\Omega,F,P}\right)$.

\par
\textbf{Assumption 2.} Suppose that the financial market involves with three tradable
assets: a bank account, a stock and a loan.

\par
\textbf{Assumption 3.} We denote the price of the bank account
 at time ~$t$ by  ~$B(t)$, such that
\begin{equation}
\label{hh1}
 dB(t) = {r}B(t)dt,~~B(0) = {B_0},~~{r} > 0,
\end{equation}
where $r$ is a constant rate of interest.

\par
\textbf{Assumption 4.} Let the price of the stock at time $t$ be $S(t)$, which  is depicted by the stochastic differential equation
(SDE).

By comparison with GBM, the O-U process is closer to the change of stock price. Here we use $S(t)$ to express the price of risk assets at $t $, which is described by O-U process
\begin{equation}
\label{hh2}
dS(t)={k}({\theta}-S(t))dt+{\sigma}dW(t),~~S(0)={S_0},
\end{equation}
where  ~$\alpha>0$, $\theta>0$ and $\sigma>0$  represent the recovery rate, the response center and the volatility, respectively.

\par
\textbf{Assumption 5.} let $R$ denote the lending rate, where $0<r<R<{\mu}$. $V(t)$ is the
pension wealth at time $t$. $V(t)$, $B(t)$ and $Y(t)$ are the total amount of money of the loan with interest, the risk-free asset and risky asset at
time $t$, respectively.

{\textbf{Definition 1.}}(Admissible strategy) If it meets the requirements as follow, investment loans are admissible.

(1)$(L(t),B(t),Y(t))$ is $\mathcal{F}_{t}$ measurable on a complete probability space;

(2)$\int_{0}^{T} {L^{2}(t)} dt < + \infty $, $\int_{0}^{T} {B^{2}(t)} dt < + \infty $, $\int_{0}^{T} {Y^{2}(t)} dt < + \infty $, a.s. $T <  \infty$;

(3)For rational investors, with interest rates higher than the deposit rate, it's impossible to choose between deposits and loans. That is, $L(t)B(t)=0 $, with $L(t) \ge 0 $, $B(t)\ge 0$ and $t\in [0,T]$.  Assume that the set of all admissible investment and loan scheme $((L(t),B(t),Y(t))$ are expressed by $\pi=\{(L(t),B(t),Y(t)):t \in [0,T]\}$.

\textbf{Assumption 6.} Define the retirement moment and the contribution rate of the enterprise annuity for  ~$T$ and $c$, separately. Where $T$ and c are the constants. Until retirement $T$, $cL(t)$ is  supplied to the pension
fund  for each period. In order to simplify the model, the total salary is set as 1 dollar, and only one insured person is studied.

\section{Model Formulation}
\subsection{Wealth process}

Let $V(t)=B(t)+Y(t)-L(t)+ct$ denote the pension wealth at time $t \in [0,T]$. The
dynamics of wealth has the following form:
\begin{align}
\label{hh3}
dV(t) =  r B(t)dt+Y(t)\frac {d S(t)}{S(t)}-RL(t)dt +cdt.
\end{align}

Based on (1) and (2),  we rewrite (3) as
\begin{align}
dV(t) =\{rX+[\frac{k(\theta-s)}{s}-r]Y+(r-R)L-rct+c\}dt+\frac{\sigma}{s}Y dt.
\end{align}

\subsection{the HJB equation}

Next, the  goal is to maximize the expected discounted utility and ultimate wealth over a limited retirement period. That is to seek the optimum investment project $Y(t)$.

Applying the stochastic control theory, we define the value function as
\begin{equation*}
H(t,s,v) = \max_{Y} E[{U(v)~|\;S(t) = s,V(t) = v}],\;\;{\rm{   }}0 < t < T,
\end{equation*}
where $U(\cdot)$ is an increasing concave utility function and satisfies
the conditions $U'(+ \infty)<0 $ and $U'(0)< +\infty $.

As described in {\citet{Soner2006Controlled}},
by the aid of It$\hat{o}$'s  formula, we have
\begin{align}
\label{hh4}
\sup_{Y}\{ H_t +& [rx+(\frac{k(\theta-s)}{s}-r)Y+(r-R)L-rct+c]H_v+\frac{k(\theta-s)}{s} H_s \\\nonumber
+& \frac{1}{2} \frac{\sigma^{2}}{s^{2}}Y^2 H_{vv}+\frac{1}{2}{\sigma^2}H_{ss}+\frac{\sigma^2}{s} YH_{vs} \}= 0,
\end{align}

and it's accompanied by a boundary condition $ H(T,s,v) = U(v) $,
where ${H_t}$,${H_s }$,${H_v }$,${H_{vv}}$,${H_{ss}}$ and ${H_{sv}}$ represent the different partial derivatives of $H(T,s,v)$.

According to $v=V(t)=B(t)+Y(t)-L(t)+ct$ and $0<r<R<{\mu}$, if $v>Y(t)+ct$, the investor will reject the loan; If $v \le Y(t)+ct$,
the investor will choose to load, but the total amount will not exceed $Y(t)+ct-v$, that is, $L^{*}(t)=Y(t)+ct-v=\max \{0,Y(t)+ct-v\}$.

From the above setting, the two situations are discussed as follows:

(1)In the case of $v \ge Y(t)+ct$, substituting $L^{*}(t)=0$  back into  (5), the HJB equation can be rewritten as
\begin{equation}
\label{hh5}
\left\{
\begin{aligned}
 &\sup\{H_t + [rv+(\frac{k(\theta-s)}{s}-r)Y-rct+c]H_v+k(\sigma-s) H_s \\
&+ \frac{1}{2}\frac{\sigma^2}{s^2}Y^{2} H_{vv}+\frac{1}{2}{\sigma^2}H_{ss}+\frac{\sigma^2}{s}YH_{vs} \}= 0 \\
&H(T,s,v) = U(v).\\
\end{aligned}
\right.
\end{equation}

(2)In the case of $v \ge Y(t)+ct$, putting $L^{*}(t)=0$  in (5), the corresponding  HJB equation can be rewritten as
\begin{equation}
\label{hh6}
\left\{
\begin{aligned}
 &\sup\{H_t + [rv+(\frac{k(\theta-s)}{s}-r)Y-rct+c]H_v+k(\theta-s) H_s \\
&+ \frac{1}{2}\frac{\sigma^2}{s^2}Y^{2} H_{vv}+\frac{1}{2}{\sigma^2}H_{ss}+\frac{\sigma^2}{s}YH_{vs} \}= 0 \\
&H(T,s,v) = U(v).\\
\end{aligned}
\right.
\end{equation}

Take the derivative of  (6) with respect to $Y$ and we have
\begin{align}
Y_{1}^{*} =&  - \frac{k(\theta-s)-rs}{\sigma^2}\frac{{H_v}}{H_{vv}} -s \frac{{{H_{vs}}}}{{{H_{vv}}}}.
\end{align}

Similarly, we can also get the efficient investment strategy of this problem(7)
\begin{align}
Y_{2}^{*} =&  -  \frac{k(\theta-s)-Rs}{\sigma^2}\frac{{H_v}}{H_{vv}} -s \frac{{{H_{vs}}}}{{{H_{vv}}}}.
\end{align}
Plugging $Y_{1}^{*}$ and  $Y_{2}^{*}$ into (6) and (7), we derive respectively
\begin{align}
\label{hh7}
{H_t} +& k(\theta-s){H_s} + \frac{1}{2}{\sigma^2}{H_{ss}} + (rx-rct+c){H_v}-\frac{{{{[k(\theta-s)  - {rs}]}^2}}}{\sigma^2}\frac{{H_v^2}}{{{H_{vv}}}}\\ \nonumber
 -&   \frac{1}{2}{\sigma^2}\frac{{H_{vs}^2}}{{{H_{vv}}}} - [k(\theta-s)  - {rs}]\frac{{{H_v}{H_{vs}}}}{{{H_{vv}}}} = 0,
\end{align}
and
\begin{align}
\label{hh8}
{H_t} +& k(\theta-s){H_s} + \frac{1}{2}{\sigma^2}{H_{ss}} + (Rx-Rct+c){H_v}-\frac{{{{[k(\theta-s)  - {Rs}]}^2}}}{\sigma^2}\frac{{H_v^2}}{{{H_{vv}}}}\\ \nonumber
 -&   \frac{1}{2}{\sigma^2}\frac{{H_{vs}^2}}{{{H_{vv}}}} - [k(\theta-s)  - {Rs}]\frac{{{H_v}{H_{vs}}}}{{{H_{vv}}}} = 0.
\end{align}

Obviously, the stochastic control problem is transformed into a NPDE. Next,  we alternate the NPDE into the linear PDE  based on the dual transformation.

\section{Model solution}
\subsection{The Legendre transform}
{\bf{Definition 2.}}Let $f:~R^n \rightarrow R $ be a convex function. Legendre
transform can be defined as follows:
\begin{equation}
\label{dfggh}
L(z) = \mathop {\sup_{x} } \{ f(x)-zx \},~~0 < t < T.
\end{equation}

Then the function $L(z)$ is called Legendre dual function of $Legendre $.

With reference to {\citet{ReviewControlled}} ,  a specific definition is proposed by
\begin{equation*}
\hat H(t,s,z) = \mathop {\sup }\limits_{v > 0} \{ H(t,s,v) - zx~|~~0 < v < \infty \},~~0 < t < T,
\end{equation*}
where $z>0$ denotes the dual variable to $v$.

The value of $v$ where this optimum is denoted by $g(t,s,z)$, so that,
\begin{equation*}
g(t,s,z) = \mathop {\inf }\limits_{v> 0} \{ v| H(t,s,v) \ge zx + \hat H(t,s,v)\},~~0 < t < T.
\end{equation*}

From the above equation, we can get
\begin{equation}
\label{hh9}
\hat H(t,s,z)= H(t,s,g) - zg,
\end{equation}
Where  $g(t,s,z) = v$ and ${H_v} = z$.

The function $\hat{H}$ is related to $g$ by
\begin{equation}
\label{hh9}
g=-{{\hat H}_z}.
\end{equation}

By differentiating (14), we achieve
\begin{equation}
\label{hh10}
{H_t} = {{\hat H}_t},~~{H_s} = {{\hat H}_s},~~{H_{vv}} =  - \frac{1}{{{{\hat H}_{zz}}}},~~{H_{ss}} = {{\hat H}_{ss}} - \frac{{\hat H_{sz}^2}}{{{{\hat H}_{zz}}}},{H_{sv}} =  - \frac{{\hat H{}_{sz}}}{{{{\hat H}_{zz}}}}.
\end{equation}

At the terminal time $T$, we define
\begin{align}
{\hat U}(z) =& \mathop {\sup }\limits_{v > 0} \{ U(z)-zv~|~~0 < v < \infty \}, \\ \nonumber
G(z)=& \mathop {\sup }\limits_{v > 0} \{ U(z)-zv~|~~0 < v < \infty \}.
\end{align}

In addition, there exists $ g(T,s,z)= (U')^{-1} $, which is a boundary condition.

Plugging (15) into (10) and (11), we derive
\begin{align}
\label{hh11}
{{\hat H}_t} +&k(\theta-s){{\hat H}_s} + \frac{1}{2}{\sigma^2}{{\hat H}_{ss}} +(rx-rct+c)z\\ \nonumber
 +&\frac{{{{[k(\theta-s)  - {r}s]}^2}{z^2}{{\hat H}_{zz}}}}{{2\sigma^2}} - [k(\theta-s)  - {r}s]z{{\hat H}_{sz}} = 0.
\end{align}
Differentiating both sides of (17) with respect to $z$, we obtain
\begin{align}
\label{hh14}
{{\hat H}_{tz}} +& [k(\theta-s)-rs]{{\hat H}_{sz}} + \frac{1}{2}{\sigma^2}{{\hat H}_{ssz}} + (rx-rct+c) + rz{g_z}+\frac{{{{[k(\theta-s)  - {r}s]}^2}{z}{{\hat H}_{zz}}}}{{\sigma^2}}  \\ \nonumber
+& \frac{{{{[k(\theta-s)  - {r}s]}^2}{z^2}{{\hat H}_{zzz}}}}{{2\sigma^2}}  - [k(\theta-s)  - {r}s]{{\hat H}_{sz}}- [k(\theta-s)  - {r}s]z{{\hat H}_{szz}} = 0 .\\\nonumber
\end{align}
Due to (14), we get
\begin{align}
\label{hh12}
v = g =&  -{\hat H}_z,~~\hat H_{tz}=-g_t,~~{\hat H}_{sz}=-g_s,~~{\hat H}_{zz}=-g_z,~~\\\nonumber
{\hat H}_{ssz}=& -g_{ss},~~\hat H_{szz}=-g_{sz},~~\hat H_{zzz}=-g_{zz}.
\end{align}

We recover (18) by using (19), and then obtain the following partial differential equation
\begin{align}
\label{hh21}
{g_t} + &[k(\theta-s)-rs]{g_s} + \frac{1}{2}{\sigma^2}{g_{ss}} -(rg-rct+c) - rz{g_z}+ \frac{{{{[k(\theta-s)  - {r}s]}^2}{z}{{g}_{z}}}}{{\sigma^2}}  \\ \nonumber
  +&\frac{{{{[k(\theta-s)  - {r}s]}^2}{z^2}{{g}_{zz}}}}{{2\sigma^2}} -  [k(\theta-s)  - {r}s]{{g}_{s}}- [k(\theta-s)  - {r}s]z{{g}_{sz}} = 0.
\end{align}

Through the dual transformation,  (10) has been transformed into a linear PDE.
Moreover, we obtain the  optimal  portfolio selection $Y_{1}^*$
\begin{align}
Y_{1}^{*} = - \frac{{[k(\theta-s)  - {r}s]sz}}{\sigma^2}g_z +sg_s.
\end{align}

\subsection{The solution under the logarithic utility function}

\noindent{\bf Theorem   4.2.1.} If the price of the risk-free asset, the price of the risk asset and the wealth process follow ~(1)-(3) respectively, the optimal portfolio  of the enterprise annuity  is specified by
according to (22)-(30)
\begin{equation}
\label{hh26}
\resizebox{.9\hsize}{!}{$Y^*(t)=\left\{
\begin{aligned}
&\frac{[k(\theta-s)  - {R}s][v-ct+cT e^{R(t-T)}]s}{\sigma^2},v \le ct+\frac{[k(\theta-s)  - {R}s][v-ct+cT e^{R(t-T)}]s}{\sigma^2} \\
&v-ct,\frac{[k(\theta-s)  - {R}s][v-ct+cT e^{R(t-T)}]s}{\sigma^2}+ct<v<
 \frac{[k(\theta-s)  - {r}s][v-ct+cT e^{r(t-T)}]s}{\sigma^2}+ct\\
&\frac{[k(\theta-s)  - {r}s][v-ct+cT e^{r(t-T)}]s}{\sigma^2},v \ge ct+\frac{[k(\theta-s)  - {r}s][v-ct+cT e^{r(t-T)}]s}{\sigma^2}.
\end{aligned}
\right.$}
\end{equation}

\noindent {\bf Proof.} In the light of the logarithmic utility function, a definition is also provided

\begin{equation*}
 U(x)= ln x ,~~~~   x > 0,~~~
\end{equation*}
Depending on the form of logarithmic utility function, we have
\begin{equation*}
g(T,s,z) = \frac{1}{z}.
\end{equation*}

In response to (20), we construct its corresponding solution
\begin{equation}
\label{hh18}
g(t,s,z) = \frac{1}{z}f({s_t}) + \varphi (t).
\end{equation}

In the meantime, we quote the boundary conditions by  $f({s_T}) = 1$ and $\varphi (T) = 0$.

Suppose it is a convex function and  we can attain
\begin{equation*}
\frac{{df(x)}}{{f(x)}} - z = 0.
\end{equation*}

Assume $x_{0}$ is the optimum point, and there is $L(z) = f({x_0}) - z{x_0}$. If $f(x) = ln x$, we have ${x_0} = \frac{1}{z}$.

As a result,
\begin{equation*}
L(z) = f(\frac{1}{z}) - 1 = ln \frac{1}{z} - 1 =  - ln z - 1
\end{equation*}
Taking the partial derivative of (22), we get
\begin{align}
\label{hh20}
g_t=& \varphi_t,~~g_s=\frac{1}{z}f+s,~~g_{z}=-\frac{1}{z^2}f~~\\\nonumber
g_{ss}=& \frac{1}{z}f_{ss},~~g_{sz}=-\frac{1}{z^2}f_s,~~g_{zz}=\frac{2}{z^3}f.
\end{align}
Substituting (23) back into (20), we obtain
\begin{equation}
\label{hh19}
{\varphi _t} + k(\theta-s)\frac{1}{z}{f_s} + \frac{\sigma^2}{2}\frac{1}{z}{f_{ss}} +rct-c-{r}\varphi  = 0.
\end{equation}

By observation, (24) can be decomposed into two equations, which is supplied to eliminate the dependence on $s$. Furthermore, since the boundary
conditions are $f({s_T}) = 1$ and $\varphi (T) = 0$, we have
\begin{equation}
\label{hh25}
\left\{
\begin{aligned}
~~&k(\theta-s)\frac{1}{z}{f_s} + \frac{\sigma^2}{2}\frac{1}{z}{f_{ss}} = 0 \\
&f({s_T})= 1,\\
\end{aligned}
\right.
\end{equation}
and
\begin{equation}
\label{hh26}
\left\{
\begin{aligned}
~~&{\varphi _t} -  {r}\varphi+rct-c  = 0 \\
&\varphi (T) = 0.\\
\end{aligned}
\right.
\end{equation}

By integrating the two equations, we derive the solution to (25)
\begin{equation*}
f({s_t}) = 1.
\end{equation*}

The corresponding solution of (26) is given by
 \begin{equation*}
 \varphi (t)  =  ct-cT e^{r(t-T)},
\end{equation*}
consequently,
 \begin{equation}
 \label{wer1}
g = \frac{1}{z}+ ct-cT e^{r(t-T)}.
 \end{equation}

Due to $g(t,s,v)=v$, we derive

 \begin{equation}
\frac{1}{z}=v-ct+cT e^{r(t-T)}.
 \end{equation}
Finally, the optimal strategy $Y_{1}^*$ can be rewritten as

\begin{align}
Y_{1}^{*} =& - \frac{{[k(\theta-s)  - {r}s]sz}}{\sigma^2}g_z +sg_s\\\nonumber
=&  \frac{{[k(\theta-s)  - {r}s]s}}{\sigma^2}\\ \nonumber
=& \frac{[k(\theta-s)  - {r}s][v-ct+cT e^{r(t-T)}]s}{\sigma^2}.
\end{align}
From the equivalence of $r$and $R$, we can get another optimal investment strategy $Y_{2}^*$
\begin{align}
\label{qw1}
Y_{2}^{*}
= \frac{[k(\theta-s)  - {R}s][v-ct+cT e^{r(t-T)}]s}{\sigma^2}.
\end{align}

The above results are discussed as follows.

(1)~If~$v \le \displaystyle \frac{[k(\theta-s)  - {r}s][v-ct+cT e^{r(t-T)}]s}{\sigma^2}+ct$ , then
\begin{align}
Y_{1}^*=\displaystyle\frac{[k(\theta-s)  - {r}s][v-ct+cT e^{r(t-T)}]s}{\sigma^2}.
\end{align}

(2)~If~$v \le (\displaystyle\frac{[k(\theta-s)  - {R}s][v-ct+cT e^{r(t-T)}]s}{\sigma^2}+ct)$ , then
\begin{align}
Y_{2}^*=\frac{[k(\theta-s)  - {R}s][v-ct+cT e^{r(t-T)}]s}{\sigma^2}.
\end{align}

(3)~If~$\frac{[k(\theta-s)  - {R}s][v-ct+cT e^{r(t-T)}]s}{\sigma^2}+ct <v <\frac{[k(\theta-s)  - {r}s][v-ct+cT e^{R(t-T)}]s}{\sigma^2}+ct$, we will proceed with two cases.

(i)With~$Y{(t)}+ct  \in [\frac{[k(\theta-s)  - {R}s][v-ct+cT e^{R(t-T)}]s}{\sigma^2}+ct,v]$, since~$L^{*}(t)=Y(t)+ct-v=\max \{0,Y(t)+ct-v\}$, then~$L^{*}(t)=0$. It means that the investment refuses to lend in this case. Let the left bracket of ~(6) be ~$\phi_{1}(Y)$.
Because ~$\phi_{1}(Y)$  is increasing with respect to ~$Y$  ~$\le[\frac{[k(\theta-s)  - {R}s][v-ct+cT e^{R(t-T)}]s}{\sigma^2},v-ct]$,
~$\phi_{1}(Y)$ attains its maximum at ~$Y^{*}(t)=v-ct$.
(ii)With~$Y{(t)}+ct  \in [v,\frac{[k(\theta-s)  - {r}s][v-ct+cT e^{r(t-T)}]s}{\sigma^2} +ct]$, since~$L^{*}(t)=Y(t)+ct-v=\max \{0,Y(t)+ct-v\}$, then ~$L^{*}(t)=v-ct$. Denote~(7) the left bracket by~$\phi_{2}(Y)$. Considering ~$\phi_{2}(Y)$
 decreases of ~$Y$ in the interval ~$[v-ct,\frac{[k(\theta-s)  - {r}s][v-ct+cT e^{r(t-T)}]s}{\sigma^2}]$, we have that ~$\phi_{2}(Y)$  reaches the maximum at ~$Y^{*}(t)=v-ct$. Hence, in the interval ~$[\frac{[k(\theta-s)  - {R}s][v-ct+cT e^{R(t-T)}]s}{\sigma^2},
 \frac{[k(\theta-s)  - {r}s][v-ct+cT e^{r(t-T)}]s}{\sigma^2}]$, $Y^*(t)=v-ct$.

In overall, the optimal investment strategy ~$Y^* (T) $can be expressed as
\begin{equation}
\label{hh26}
\resizebox{.9\hsize}{!}{$Y^*(t)=\left\{
\begin{aligned}
&\frac{[k(\theta-s)  - {R}s][v-ct+cT e^{R(t-T)}]s}{\sigma^2},v \le ct+\frac{[k(\theta-s)  - {R}s][v-ct+cT e^{R(t-T)}]s}{\sigma^2} \\
&v-ct,\frac{[k(\theta-s)  - {R}s][v-ct+cT e^{R(t-T)}]s}{\sigma^2}+ct<v<
 \frac{[k(\theta-s)  - {r}s][v-ct+cT e^{r(t-T)}]s}{\sigma^2}+ct\\
&\frac{[k(\theta-s)  - {r}s][v-ct+cT e^{r(t-T)}]s}{\sigma^2},v \ge ct+\frac{[k(\theta-s)  - {r}s][v-ct+cT e^{r(t-T)}]s}{\sigma^2}.
\end{aligned}
\right.$}
\end{equation}

\noindent{\bf  Theorem 4.2.2.} If the price of the risk-free asset, the price of the risk asset and the wealth process follow ~(1)-(3) respectively, the expected maximum utility of the enterprise annuity for problem ~(10) and (11) is

\noindent {(1)In the case of $v \ge Y(t)+ct$,}
\begin{equation*}
H_1=ln (v-ct+cT e^{r(t-T)}).
\end{equation*}
\noindent{(2)In the case of $v<Y(t)+ct$,}
\begin{equation*}
 H_2=ln (v-ct+cT e^{R(t-T)}).
 \end{equation*}
 \noindent {\bf Proof. }We first prove the first case.
Combining(27) with $-\hat H_{z}(t,s,v)=g $, we have
\begin{equation}
\label{hh22234}
 \hat H_{z}=-\frac{1}{z}-ct+cT e^{r(t-T)}.
\end{equation}
From (34), integrating yields
\begin{equation*}
\label{hh202}
 \hat H=-lnz+-ctz+czT e^{r(t-T)}+m,
\end{equation*}
Where $m $ is a constant.

Taking into account $\hat H= H - zg$ and the terminal condition $m=-1$, we obtain
\begin{equation*}
 H_1=ln (v-ct+cT e^{r(t-T)}).
\end{equation*}
By the same token, we derive
\begin{equation*}
 H_2=ln (v-ct+cT e^{R(t-T)}).
\end{equation*}

\section{ Numerical analysis}
Based on these simulation results, we
provide some economic explanations and discuss the behavioral
features related to loss aversion, and contribution rate. We take the initial  time $t=5$,  and the investor will retire at $T=20$. In the financial market,  other parameters are  ${r}=0.03$, ${R}=0.06$, ${\sigma_1}=0.005$ and ${c}=0.2$.

In Fig. 1 and Fig. 2, the volatility $\sigma$ on the optimal investment strategy $Y^*(t)$ is taken into account.  Assume that the wealth value  is 500, 1200 at time t, respectively, and the volatility
$\sigma$ varies at $[1,2.4]$. If $v=500$, then $v \le ct+\frac{[k(\theta-s)  - {R}s][v-ct+cT e^{R(t-T)}]s}{\sigma^2}$.
If $\sigma$ varies in the range $[1,2.4]$ and $v=1200$, then $v \ge ct+\frac{[k(\theta-s)  - {r}s][v-ct+cT e^{r(t-T)}]s}{\sigma^2}$.
¡¢

\def\figurename{\kaishu ͼ}
\begin{center}
\centering
\includegraphics[width=0.6\textwidth]{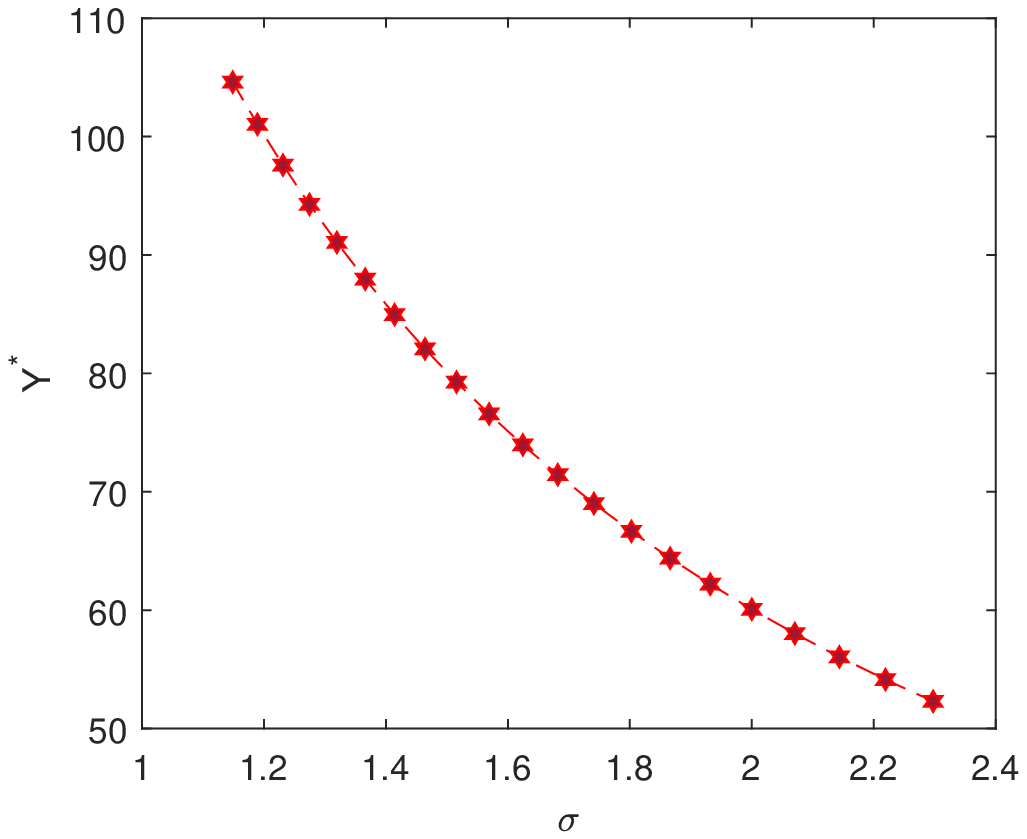}
\end{center}
\begin{center}Fig. 1. Effect of parameter $\sigma $ on $Y^*$ when $v=500$
\end{center}

\def\figurename{\kaishu ͼ}
\begin{center}
\centering
\includegraphics[width=0.6\textwidth]{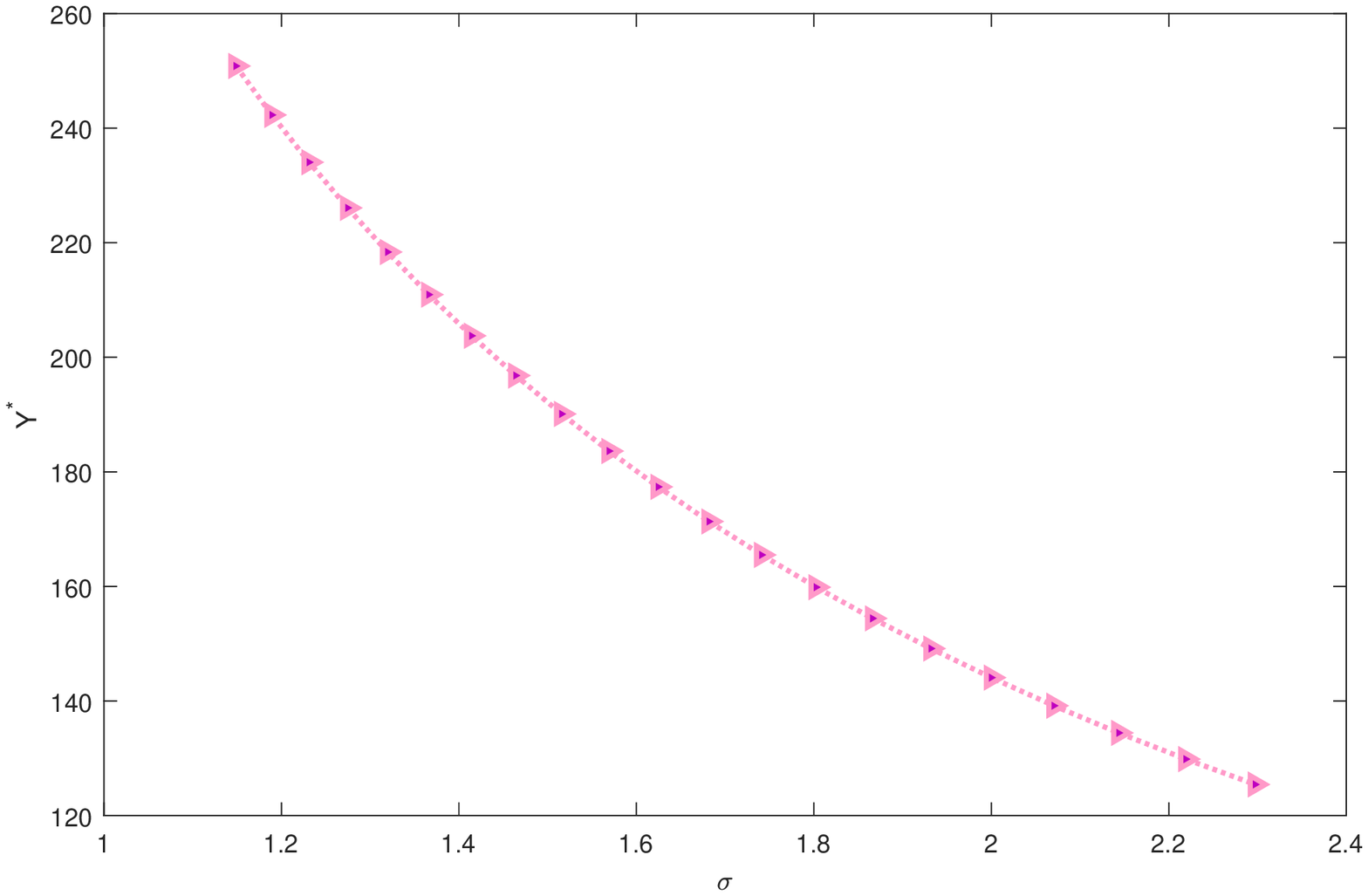}
\end{center}
\begin{center}Fig.2. Effect of parameter $\sigma $ on $Y^*$ when $v=1200$
\end{center}

Fig.2 presents the impacts of $\sigma $ on $Y^*$ invested in the risky asset. We can see, from Fig.1, $Y^*$ reduces while  $\sigma $ grows. In an economic sense, the more drastic the stock price varies, the more uncertain the market is.
Investors are afraid to take risks and will engage in conservative risk-averse behaviors, that is, they will increase their investment in risk-free assets.

\def\figurename{\kaishu ͼ}
\begin{center}
\centering
\includegraphics[width=0.6\textwidth]{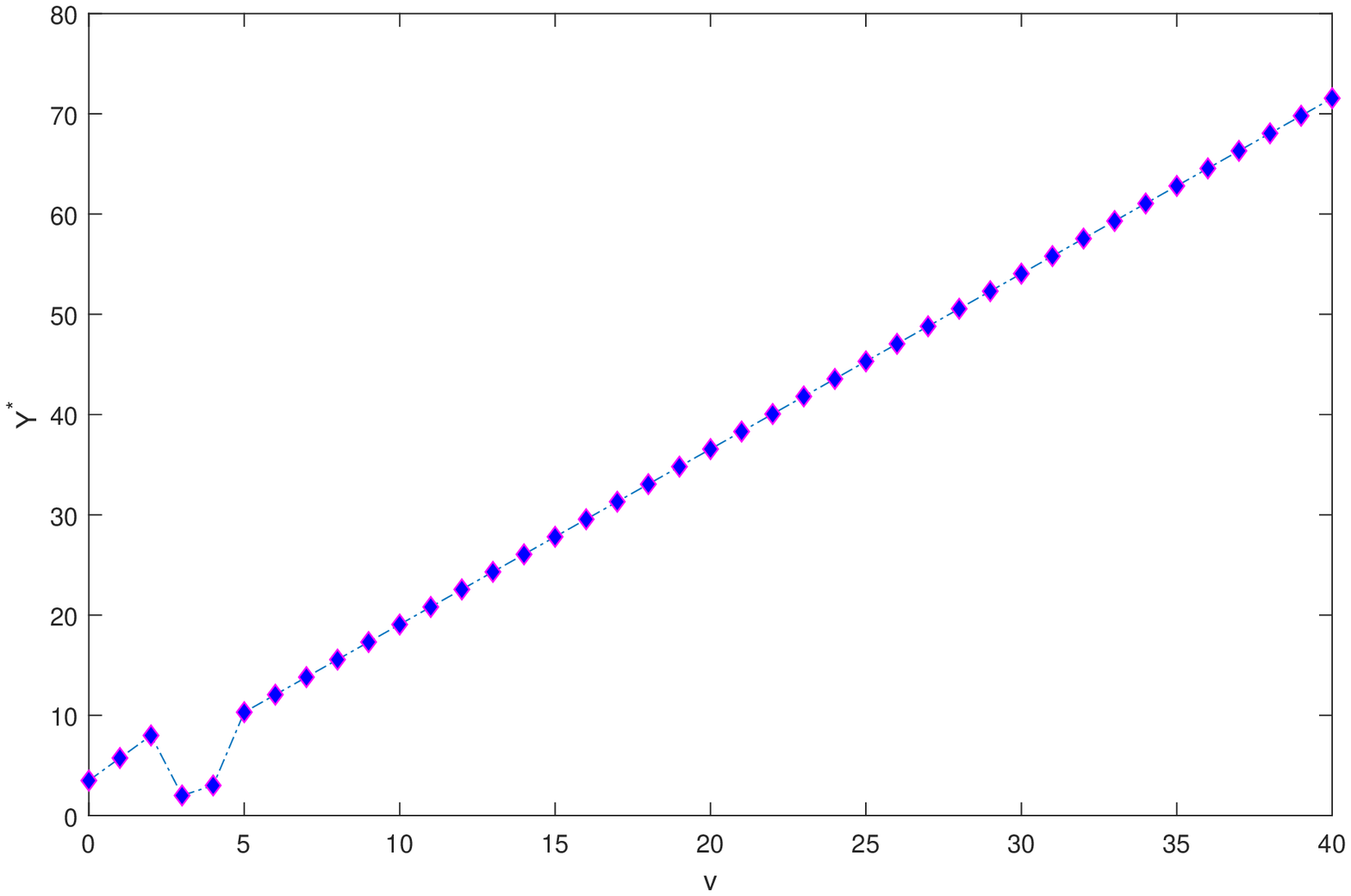}
\end{center}
\begin{center}Fig.3. Effect of parameter $v$ on $Y^*$ when $R>r$
\end{center}

Fig. 3 shows $v$ on   $Y^*$ invested in the risky asset. We adopt the instantaneous volatility ~$\sigma=0.2$.
From Fig. 3, we may find that $Y^*$ increases with the initial wealth $v$. This
can be explained by the fact that  employees become richer, they become more capable of taking risks. Therefore, the pension manager tend to spend money on risky assets to get more return.

\def\figurename{\kaishu ͼ}
\begin{center}
\centering
\includegraphics[width=0.6\textwidth]{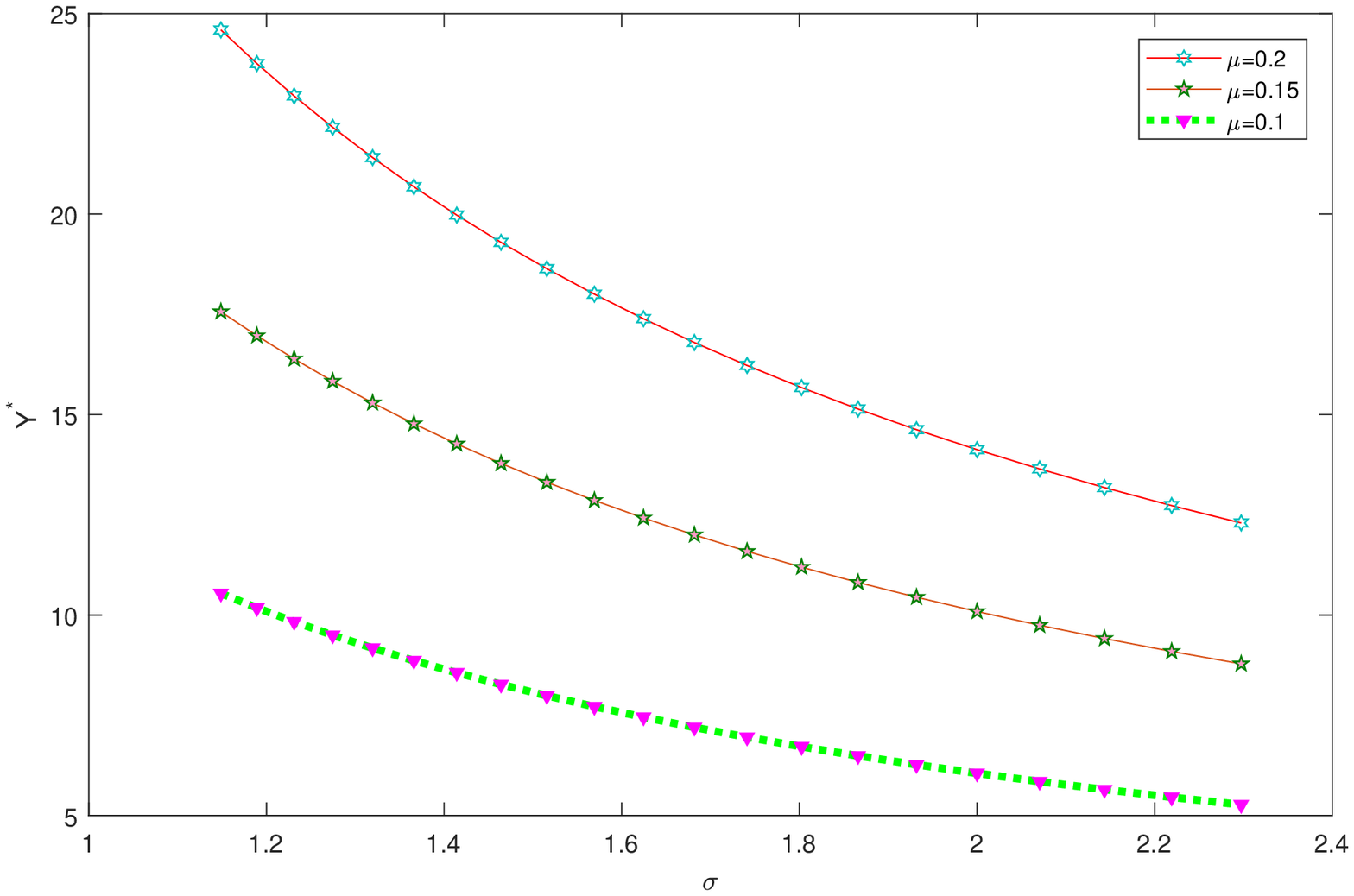}
\end{center}
\begin{center}Fig. 4. Effects of parameters $\sigma $ and $\mu$ on $Y^*$
\end{center}

Fig. 4 displays  ~$\sigma $ and  $\mu$  on the robust optimal investment strategy.
As shown in Fig. 4,  ~$Y^*$  decreases with  regard to ~$\sigma $. A higher ~$\sigma$ leads
to a larger expected drop in volatility and an increased probability
of a large adverse movement in the risky asset¡¯s price. In addition, under the elasticity coefficient ~$\sigma $  is fixed, when $\mu$ is raising, $Y^*$  also rises.
This is because that an increase in the expected instantaneous rate makes the member improve his ability to resist risk,
and hence she invests more in  the stock.

\def\figurename{\kaishu ͼ}
\begin{center}
\centering
\includegraphics[width=0.6\textwidth]{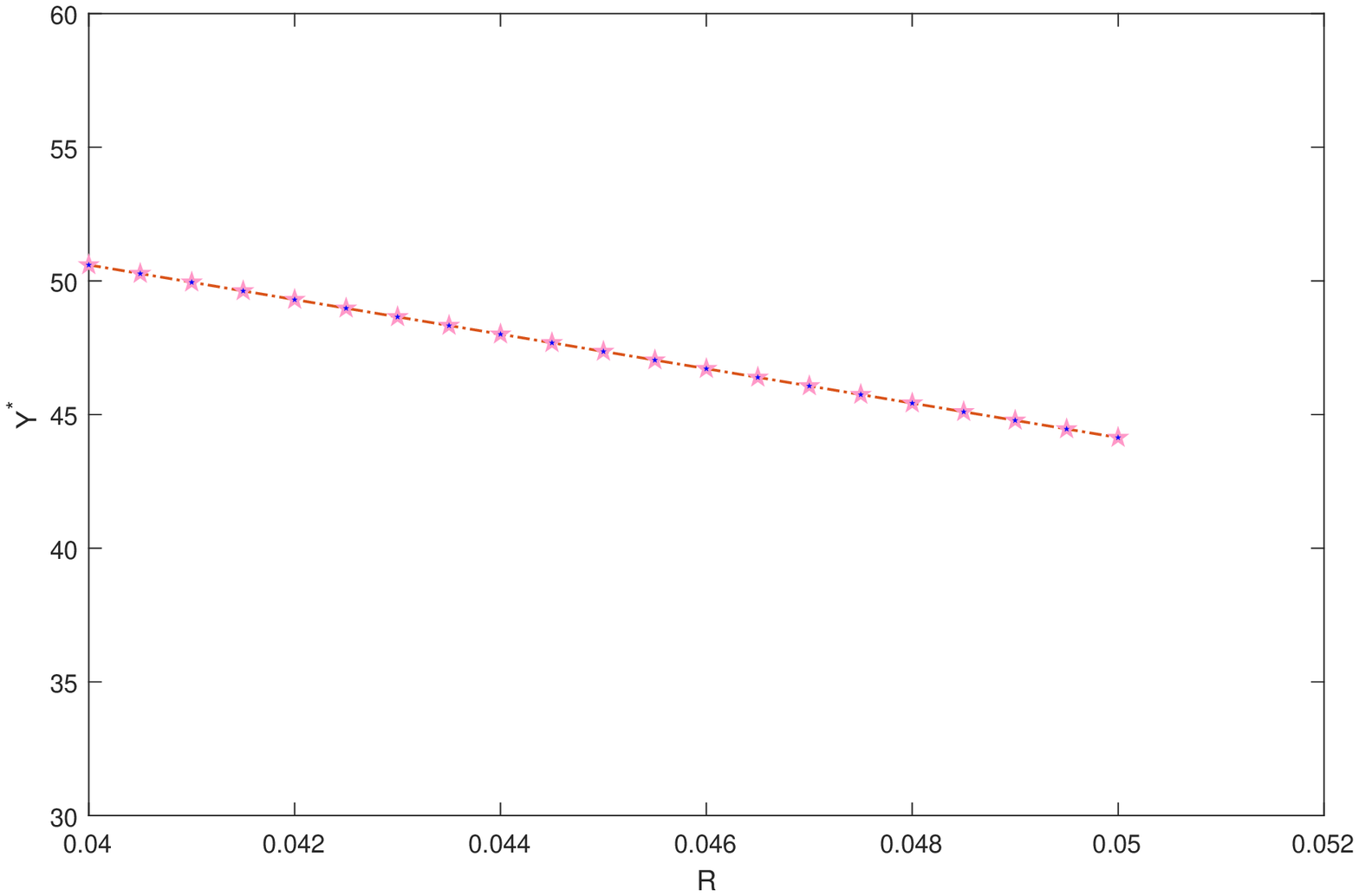}
\end{center}
\begin{center}Fig. 5. Effect of parameter $R$ on $Y^*$
\end{center}

Fig. 5 displays that as the lending rate $R$ increases, the proportion of wealth
invested in the stock becomes larger. With the increase of $R$ , there is more risk in the market and  it takes a lot of time to
 invest.
As a result, the manager will reduce the amount of money on risk assets which  is also in line with the economic market.

\section{Conclusion}
Optimal portfolio has always been the core of financial market research.
We do the research about problem
with the the O-U process under $CRRA$ utility.
With the help of dynamic programming principle and dual transform method, the closed-form of optimal asset allocation strategy is obtained.
Finally, ~MATLAB software is used for programming. More importantly,
we do an analysis of the volatility of the stocks, the initial wealth, the elasticity coefficient and the lending rate on the investment behavior.

~~~~

\noindent {\bf Acknowledgements}

~~~~

The authors are very grateful to two anonymous referees for their valuable and constructive advice on the thesis.

\footnotesize{\bibliography{mybibfile}

\biboptions{authoryear}


\bibliographystyle{elsarticle-harv}%

\end{document}